# A simple theory of molecular organization in fullerene containing liquid crystals.

S. D. Peroukidis, A. G. Vanakaras(*) and D. J. Photinos
*University of Patras, Department of Materials Science, Patras 26504, GR*

**Abstract**

Systematic efforts to synthesise fullerene containing LCs have produced a variety of successful model compounds. We present a simple molecular theory relating the self-organisation observed in these systems to their molecular structure. The interactions are modelled by dividing each molecule into a number of sub-molecular blocks to which specific interactions are assigned. Three types of blocks are introduced, corresponding to fullerene units, mesogenic units, and non-mesogenic linkage units. The blocks are constrained to move on a rectangular 3-dimensional lattice and molecular flexibility is allowed by retaining a number of representative conformations within the block representation of the molecule. Calculations are presented for a variety of molecular architectures including twin mesogenic branch mono-adducts of $C_{60}$, twin dendro-mesogenic branch mono-adducts and conical (badminton shuttlecock) multi-adducts of $C_{60}$. In spite of its many simplifications, the theory accounts remarkably well for the phase behaviour of these systems.

## I.     Introduction

The possibility of controllable formation of ordered molecular assemblies of fullerenes is of fundamental importance for the use of these allotropes in applications[1-3]. However, the strong interactions among fullerene molecules normally lead to the formation of aggregates and therefore addends are used in order to modify favourably these interactions. The covalent linkage of fullerenes to liquid crystal (LC) forming molecules (mesogens) offers a way to self organised structures in which the fullerenes form ordered molecular assemblies [4]. Viewed from the LC perspective, the fullerination of mesogens establishes a new direction in the design of mesomorphic functional materials, given the peculiar photo-and electro-chemical properties of the fullerene molecule.

Starting in 1996 with the work of Chuard and Deschenaux [5], systematic efforts to synthesise fullerene containing LCs have produced a variety of successful model compounds. These compounds are formed by the covalent linkage of one or more mesogenic units, typically of the calamitic (rod-like) type, at one or more sites of the fullerene frame. Depending on the structure of the mesogenic part and the topology of the fullerene–mesogen linkage, these compounds can be grouped into the following types, with the respective architectures illustrated in figure 1.

(a) Twin mesogenic branch mono-adducts [4-9] of $C_{60}$ in which two branches are attached, typically via a methanofullerene connecting group. The branches start out with a flexible alkyl spacer and terminate with a rod-like mesogenic unit (figure 1a). Compounds of this type have been reported to form exclusively smectic-*A* mesophases.

(b) Twin dendritic branch mono-adducts [6-8,10-12] of $C_{60}$, with two branches, as in (a), except that each branch is linked to a whole dendrimer (figure 1b), the branches of which are functionalised with mesogenic units (dendro-mesogen). The dominant LC phase exhibited by these super-mesogens is the smectic-*A* mesophase, although the possibility of formation of nematic phases, in addition to the smectic-*A* phase, has been reported for certain low-generation denro-mesogen addends [7,12].
(c) Single dendro-mesogenic branch mono-adducts [7,8,13,14] of $C_{60}$. These are modifications of (b) in which one of the two branches terminates in a dendro-mesogenic unit and the other has a non-dendritic terminus (figure 1c). These compounds are reported to exhibit smectic-*A* mesophases.
(d) Dendro-mesogenic bis-adducts [15] of $C_{60}$ (figure 1d) with two dendro-mesogenic branches as in (b). The structure of the mesophase formed by these compounds has not been identified conclusively.
(e) Mesogenic branch hexa-adducts [16] of $C_{60}$. Six pairs of twin branches, each bearing a terminal mesogenic part, are attached via six methanofullerene connecting groups (figure 1e). These super-mesogens form smectic-*A* phases.
(f) Conical (badminton shuttlecock) multi-adducts [17-19] of $C_{60}$ (figure 1f). The conical surface is formed by the direct attachment of (five) mesogenic units via single bonds to carbon sites of the fullerene, which thus becomes the apex of the cone. Hexagonal- and nematic-columnar mesophases have been reported for these super-mesogens.

The variety of fullerene containing LC compounds is already broad enough to permit the deduction of certain trends and possibly empirical design rules and also to provide testing grounds for a molecular theory. The purpose of this paper is to introduce a simple molecular theory relating the self-organisation exhibited by these systems to their molecular structure.

The molecular interactions are modeled in a modular fashion: the fullerene containing mesogen is divided into a number of chemically distinct sub-molecular components (modules) to which specific interactions are assigned. The total interaction of the molecular ensemble is then built up as a combination of interactions between all the possible pairs of modules within the ensemble. To simplify the computational aspects of the theory, the modular subdivision of the molecules is done rather coarsely, distinguishing only three types of sub-molecular components: fullerene units, mesogenic units, and non-mesogenic units (flexible spacers and linkage groups). For the same reasons of simplicity, the molecules are taken to move on a rectangular 3-dimensional lattice.

In its present primitive form, the theory attempts to describe in a unified way only the basic trends observed in the self-organisation of these systems. It is not intended to provide a quantitative description or to account for peculiarities. Indeed, it is well known from the study of many conventional "simple" LCs that apparently small changes in the molecular structure could bring about dramatic changes in the self-organisation. The treatment of analogous situations in fullerenated LCs is clearly beyond the reach of the present form of the theory.

The modelling of the molecular conformations and interactions are described in section II. The results of calculations for various architectures of fullerenated LCs are presented and their significance is discussed in section III. The conclusions from this work are stated in section IV.

**II. Molecular modelling on a cubic lattice**
The statistical mechanics machinery of the present theory is based on the interconverting conformer formulation of the variational cluster method presented in [20]. In this section we describe the application of this approach to study the phase behaviour of fullerene containing super-mesogens of various architectures. As detailed in ref [20], the necessary ingredients for the implementation of this approach are:

(i) A number of discrete molecular states or "shapes" representing the basic types of molecular conformations (the "conformers").

(ii) The assignment of an intrinsic free energy $\varepsilon_S$ to each of these states. Equivalently an intrinsic probability $P_S^0$ is assigned to each state $S$, giving the probability of finding an *isolated* molecule in that state.

(iii) The coarse-grained subdivision of the molecules into blocks with specified interactions. The overall molecular interactions are then described by a modular pair-potential combing the interactions among all possible inter-molecular pairs of blocks.

These ingredients are molecular characteristics that can be obtained to the desired detail by simple molecular mechanics calculations on the atomistic scale. However, as in this work the focus is on the qualitative picture of mesophase description rather than on the quantitative accuracy, a minimal number of representative molecular states will introduced, with rough estimates for their intrinsic probabilities, and a rather coarse molecular subdivision will be used together with a very simple parameterisation of the sub-molecular block-block interactions. Thus, based on the presence of the fullerene molecule in the systems presented in figure 1, we have chosen to model their chemical structures on the length scale of the diameter of a single fullerene. An analogous discretisation, on that length scale, is imposed on the molecular motions in space: the molecules are taken to move on a rectangular 3-dimensional lattice with unit cell dimensions equal to a fullerene diameter. The coarse grained molecular shapes are then tailored using building blocks of size equal to the unit cell of the lattice space. In all cases considered here the conformers consist of a building block corresponding to the fullerene unit and of several blocks connected to build up the grafted addends according to the molecular architecture of the conformer. Block representations for selected molecular architectures from figure 1 are depicted in figures 2-4.

Once the molecular shapes have been built the interaction potential $U_{I,J}$ between molecules $I$ and $J$ is obtained in a modular fashion [20] by combining interaction contributions among intermolecular pairs of blocks,

$$U_{I,J} = \sum_{b_I,b_J} \left( u^{(0)}_{b_I,b_J} \delta(r_{b_I,b_J}) + u^{(1)}_{b_I,b_J} \delta(|r_{b_I,b_J}|-1) \right) , \qquad (1)$$

where $r_{b_I,b_J}$ denotes the distance between blocks $b_I$, $b_J$, and $u^{(0)}_{b_I,b_J}$, $u^{(1)}_{b_I,b_J}$ stand for the interaction potentials for an pair of blocks occupying the same or adjacent lattice sites respectively. In what follows the block index $b$ can be either $f$ or $m$ or $l$, denoting respectively fullerene units, mesogenic units, and non-mesogenic units (flexible spacers and linkage groups). We assume the following general form for the intermolecular block-block interaction potential $u^{(n)}_{b_I,b_J}$, $n = 0,1$

$$u^{(n)}_{b,b'} = q^{(n)}_{b,b'} + w^{(n)}_{b,b'} P_2(\mathbf{e}_b \cdot \mathbf{e}_{b'}) \qquad (2)$$

with $q^{(n)}_{b,b'}$ and $w^{(n)}_{b,b'}$ being phenomenological parameters for the strength of the interactions among blocks $b$ and $b'$ when they occupy the same, $n = 0$, or adjacent, $n = 1$, cells.

The $q^{(n)}_{b,b'}$ terms define the strength of the non-directional (isotropic) part of the block-block interactions. It should be noted, however, that these terms alone are sufficient to generate a directionality in the *overall interaction* among the molecules if the (non-directional) blocks are connected together in a directional manner (as it is the case in all the block representations of the conformers in figures 2-4). The terms $w_{b,b'}^{(n)}$ in equation (2) allow for the inclusion of intrinsic directionality in the block-block interactions. This would account for the anisotropic interactions among mesogenic units, in the case where blocks $b$ and $b'$ correspond to such units. The directionality is conveyed simply by the Legendre polynomial of second order, $P_2(\mathbf{e}_b \cdot \mathbf{e}_{b'})$, where $\mathbf{e}_b$ and $\mathbf{e}_{b'}$ are unit vectors defining the long axes of the mesogenic units. In the block constructions used here (figures 2-4), these axes coincide with the directions of the branches to which the mesogenic units belong. Polar interactions among blocks have been ignored but they could readily be accounted for in more elaborate calculations by including $P_1(\mathbf{e}_b \cdot \mathbf{e}_{b'})$ terms in equation (2). Their omission from the present calculations is done for reasons of simplicity and does not necessarily imply that their effects are negligible. In fact the molecular structures of most of the compounds considered in this study include segments with strong electric dipole moments and such moments are known to affect significantly the relative stability of liquid crystalline phases of common mesogens [21,22].

Next, we describe in some detail the conformer structure and intrinsic probabilities as well as the block structure and interactions for three generic architectures of fullerene containing supermesogens.

**II.1. Twin mesogenic branch of $C_{60}$ (TMB-$C_{60}$).** These consist of two identical branches grafted on the surface of a $C_{60}$ fullerene, figure 1a, with each branch bearing one mesogenic unit. Clearly, the number of accessible conformations of such systems is quite large due to the flexibility of the spacers connecting the $C_{60}$ moiety with the mesogenic units. Furthermore, molecular-mechanics calculations with widely accepted empirical force fields [23] indicate that conformations of completely different overall shapes may not differ substantially in their energies. In addition, molecular states with the mesogenic units on the same side of the $C_{60}$ and mutually parallel, despite some energetic gain due to Van der Walls attractions between the mesogenic units (compared to states with the mesogenic units far apart), are achieved for a limited number of energetically accessible conformations of the flexible spacers and thus become entropically disfavoured.

Based on these considerations we model the TMB-$C_{60}$ systems by grouping the molecular conformations in three distinct coarse-grained molecular shapes. Each shape consists of a single fullerene block and two branches. With building blocks of the fullerene size, the actual length of each branch is roughly three times the fullerene diameter. Consequently, each branch consists of three linearly attached building blocks of which the one closest to the fullerene is partially occupied by the flexible spacer and the next two occupied by the mesogenic unit. In the first of the shapes, to be referred to as the extended antiparallel (EA) conformer (denoted by EA), the two branches extend in opposite directions rendering the conformer cylindrical symmetric (strictly, four-fold symmetric on the cubic lattice) and apolar (figure 2a). In the second, the "L"-shaped conformer (denoted by L), the two branches are perpendicular one to the other and the molecular shape lacks both apolarity and cylindrical symmetry (figure 2b). In the third conformer, to be referred to as the folded parallel (FP) conformer (denoted by FP) both branches share the same space (figure 2c) and the molecular block representation is cylindrically symmetric and polar.

For this class of systems the grafted branches on the fullerene are not very bulky. Therefore we assume that any two submolecular blocks interact only when they occupy the same lattice site. Consequently $u_{b,b'}^{(1)}$ is taken to vanish for any pair of blocks. For the parameterization of the $u_{b,b'}^{(0)}$ potential we assume that the lattice sites containing fullerene building blocks are not permitted to be occupied by any other block, i.e. fullerene blocks are impenetrable both, to other fullerene blocks and to blocks corresponding to branches. Furthermore, the building blocks of the branches are assumed exert on one another soft repulsions without any directionality. In the parameterisation of equation (2), these assumptions correspond to $w_{b,b'}^{(0)} = 0$ for any intermolecular pair of blocks $b,b'$, $q_{f,b}^{(0)} = \infty$ for $b$ either $f$ or $m$ or $l$, and $q_{l,l}^{(0)} = q_{m,m}^{(0)} = q_{l,m}^{(0)} = q_0 > 0$. The implications on the molecular organisation of the TMB-$C_{60}$ upon the inclusion of directional interactions between branches are also considered in some of our calculations by allowing for $w_{b,b'}^{(0)} \neq 0$.

The intrinsic free energies of the conformers with linear antiparallel or perpendicular branches are assumed to be equal, $\varepsilon_{EA} = \varepsilon_L = 0$. For the intrinsic free energy $\varepsilon_{FP}$ of the remaining conformer FP, with both branches extended in the same direction, we consider the three possibilities $\varepsilon_{FP} > 0$, $\varepsilon_{FP} = 0$ or $\varepsilon_{FP} < 0$ corresponding to this conformer being intrinsically less-, equally- or more-probable than the other two conformers.

**II.2.    Twin dendro-mesogenic branch of $C_{60}$ (TDB-$C_{60}$).**    This is the case of liquid crystalline dendrons attached to the $C_{60}$ surface giving rise to molecular architectures similar to those in figures 1b and 1d and their higher generation counterparts. The chemical structure of a second generation mesogenic dendrimer grafted at a single point on the fullerene surface [7,8] is depicted in its fully extended conformation (the two dendritic branches extending in opposite directions) in figure 3a. In accordance with molecular mechanics calculations, conformations with both dendritic branches in the same direction (not shown in the figure) are also possible. Therefore we assume that the molecular conformations can be grouped in two dominant molecular shapes with the dendritic units extending either on the same side (folded parallel, FP, conformer) or to opposite sides (extended antiparallel, EA, conformer) and we have built these conformers according to the shapes shown in figures 3b-3c. At this level of resolution, the primary difference from the structures of figure 2 is that the arms grafted to the fullerene are bulkier. It should be noted that different tones of shading have been used in figure 3 to distinguish between blocks of different content. We have used the darker shading for the addends of the FP conformer to indicate that, in this case, the addend-blocks of both branches share the same space, having therefore twice the density of the corresponding blocks of EA conformer.

Calculations were performed initially with the inclusion of a third, "L"-shaped, conformer. These calculations showed, however, that the inclusion of this conformer does not alter significantly the phase behaviour of the system. The parameterisation of the block-block interactions when the blocks occupy the same lattice site is similar to the corresponding parameterization for the TMB-$C_{60}$ blocks. In order to convey the bulkier nature of the dendritic addends of the TDB-$C_{60}$ we have assumed that the addend-blocks repel softly each other even when they are in adjacent cells. The form of the interaction potential we have used for the calculations for the block model of TDB-$C_{60}$ is summarized in table I.

It should be noted here that parameterisation is expected to break down for grafted dendro-mesogens of generation higher than the third. In that case, the fullerene size becomes very small compared to the rather bulky branches which would thus cover the fullerenes completely, therefore preventing any direct fullerene-fullerene interactions.

**II.3.    Conical super-mesogens with a fullerene apex (CSM-$C_{60}$).**    These are shown in figure 4a together with the respective block structure used in the present modelling. Here we assume two kinds of molecular building blocks, the fullerene blocks ($f$-blocks) and the blocks that correspond to the grafted addends (*m*-blocks). A single conformer is assumed for these systems. This renders the block representation of the supermesogen rigid. Such representation is in line with the chemical structures of the CSM-$C_{60}$ where five aromatic groups are attached around a pentagon of the fullerene molecule [17-19] thus forming an essentially rigid "nano-shuttlecock". Certainly, the presence of aliphatic end-chains on the grafted mesogenic groups of the real systems introduces some molecular flexibility which, however, entails only minor deviations from the dominant hollow-cone molecular shape.

For a broader assessment of the significance of the results presented in the next section, it is worth noting that, first, the theoretical framework outlined here is not restricted to fullerene containing LCs nor is the lattice representation an inherently restricting feature of the theory; it merely reduces the computational effort without seriously distorting the essence of the molecular description. Secondly, the assignment of the *inter*-block interactions adopted here is based mainly on intuition. However, on taking into account the detailed structure of chemical units that fill the building blocks, the use of simple molecular mechanics calculations can provide more accurate estimates of the *inter*-block interactions.

### III. Results and discussion

Here we present results on the phase behaviour of the three types of model structures introduced in the previous section. Starting with the twin mesogenic branch (TMB) systems, different situations are explored within the parameterisation adopted for the block-block interactions (strength parameter $q_0$) and the intrinsic probabilities of the molecular conformations (parameter $\varepsilon_{FP}$). For all the combinations of $q_0$ and $\varepsilon_{FP}$ studied, the mesophases formed by these systems are nematic and orthogonal smectic phases of different layer structures.

Due to the possibility of inter-conversions between molecular states of significantly different molecular length, the smectic polymorphism found in these systems has some notable differences from the polymorphism of conventional smectic compounds where a single molecular length is dominant. It is known that strongly polar rodlike molecules give rise to a rich polymorphism of orthogonal layered mesophases [22,24-27]. Thus, distinctions of the smectic-*A* (Sm*A*) phases into Sm$A_1$, Sm$A_d$ and Sm$A_2$ phases are introduced, where the index 1, *d*, 2, indicates that the wavelength of the periodic density modulation is, respectively, one, *d* (1<*d*<2) or two times the molecular length. Moreover, incommensurate Sm$A_i$ phases have been reported [26,27]. These phases are characterized by spatial modulations along the nematic director with wavelengths $\ell$ and $\ell'$ of irrational $\ell/\ell'$ ratio. This classification of the smectic-*A* phases assumes the existence of a single molecular length (or at least a narrowly defined range of molecular lengths). Clearly this is not the case for the TMB systems of figure 2 where the three dominant states of the molecule are vastly different in their geometrical characteristics. However, for the purpose of the present study we have labeled the smectic phases in close analogy to the widely accepted nomenclature of the conventional smectic compounds, by defining the "molecular length" to be the sum of the grafted branch length and the fullerene diameter. According to this choice of molecular length measure, we denote by Sm$A_{d1}$ highly interdigitated smectic phases where the fullerenes within a smectic layer form a single sub-layer of thickness equal to the fullerene diameter (fullerene monolayer). Smectic-*A* structures with structure similar to Sm$A_{d1}$ but with the thickness of the fullerene-rich sub-layers being twice the fullerene diameter are denoted as Sm$A_{d2}$ (fullerene bi-layer within a single smectic layer). Finally, incommensurate smectic phases with structures corresponding to a superposition of $SmA_{d1}$ and $SmA_{d1}$ are denoted as $SmA_{di}$.

The phase behaviour of the TMB systems, adopting the above nomenclature for the layered phases, is summarized in figures 5a-5d where we plot pressure *vs* reciprocal temperature phase diagrams for various values of the scaled intrinsic free energy difference $\varepsilon_{FP}/q_0$. In all cases, the systems exhibit an isotropic phase (*I*), different orthogonal smectic phases and a low temperature uniaxial nematic phase (*N*) which however, is stable over a rather narrow region of the phase diagram (not shown at all within the plotted temperature range in figure 5a). The orthogonal smectic phases differ in the way the fullerenes organize within the smectic layers and also in the degree of interdigitation of the arms in adjacent smectic layers. Specifically, in the high-temperature, high-pressure smectic phases the fullerenes are arranged on a single sub-layer (fullerene monolayer, $SmA_{d1}$) while in the lower temperature smectic phases the fullerenes occupy two successive sublayers (fullerene bi-layer, $SmA_{d2}$).

All the phase transitions are of first order. The $SmA_{d1}$-$SmA_{d2}$ transitions and the isotropic to nematic are weaker than the isotropic to smectic and nematic to smectic. The nematic phase, for all the studied systems, appears only at low temperatures where the conformers become practically impenetrable and their self-organization is determined primarily by the overall shape anisotropy.

For $\varepsilon_{FP}/q_0 = -3$, namely when folded, FP, conformers are intrinsically much more probable than the extended conformers (EA and L), the $SmA_{d2}$ molecular organisation dominates the layered mesophases except for a small window at high pressures-low temperatures where the $SmA_{d1}$ phase is more stable, figure 5a. In this case the nematic phase appears only at very low temperatures. Increasing the free energy $\varepsilon_{FP}$ but still keeping it negative $\varepsilon_{FP}/q_0 = -1$ the overall topology of the phase diagram (figure 5b), is the same but with larger $SmA_{d1}$ phase region. The tendency towards the stabilization of a $SmA_{d1}$ type of molecular organization is further strengthened when $\varepsilon_{FP}/q_0 = 1$ (figure 5c). In that case, with the folded parallel conformer having higher intrinsic free-energy (lower probability) than the extended anti-parallel conformer, the $SmA_{d1}$ window extends to higher temperatures. For the cases mentioned above there are three possibilities of phase sequences on decreasing the temperature at constant pressure: (a) the high pressure phase sequence *I*-$SmA_{d1}$-$SmA_{d2}$ (b) the intermediate pressure sequence *I*-$SmA_{d2}$ and (c) the low pressure sequence *I*-*N*-$SmA_{d2}$

The topology of the phase diagram and subsequently the possible phase sequences change dramatically when $\varepsilon_{FP}/q_0 = 3$ (figure 5d). In that case the region of stability for the $SmA_{d1}$, is greatly enhanced covering, for moderate pressures, the whole range of temperatures. Thus the phase sequences *I*-$SmA_{d1}$ and *I*-*N*-$SmA_{d1}$, not observed before, become possible when the molecular conformers with their branches separated (EA or L conformers) are given high intrinsic probabilities.

Near the smectic-smectic phase transitions the spacing of the layered phases increases significantly. This is found not to be due to a weakening of the molecular interdigitation but rather to the formation of thermodynamicaly stable incommensurate smectic phases with the layers divided into sublayers exhibiting both $SmA_{d1}$ and $SmA_{d2}$ type of molecular organisation. These intermediate phases are stable over very narrow ranges, appearing in a nearly continuous succession between $SmA_{d1}$ and $SmA_{d2}$.

The molecular organisation in the smectic phases and the relative population of the conformers are dictated mainly by microsegregation. This is demonstrated in figure 6 where we have plotted the bulk probability of the three molecular shapes as function of pressure at a fixed value of the scaled temperature $q_0/k_B T = 0.20$ for the system with $\varepsilon_{FP}/q_0 = -1$. On the same diagram, the straight lines correspond to the intrinsic probability of the EA conformer and the L-shaped conformer (solid line) and of the FP conformer (dashed line). It is clear from this plot that the L conformer becomes substantially less probable in all ordered phases while it is equally probable with the EA conformer in the isotropic phase. The marked changes of the bulk probabilities across the phase transitions reveal the conformational nature of the transitions.

We have repeated the calculations retaining only the two linear conformers (EA and FP) and omitting the L conformer. In this case, apart from a relatively weak stabilisation of the ordered phases with respect to the isotropic phase, the phase diagrams are qualitatively the same with those obtained by retaining all three conformers. This demonstrates that molecular shapes whose symmetries deviate significantly from the symmetries of the phase have minor influence on the molecular organisation in the ordered phases since they are strongly suppressed within the bulk phase even though their intrinsic probability is comparable to that of the dominant shapes.

We have also investigated the TMB systems in the presence of intrinsically directional block-block interactions for the end blocks, which correspond to mesogenic units. We have considered orientational interactions whose relative strength $r \equiv w_{m,m}^{(0)}/q_{m,m}^{(0)}$ with respect to the isotropic interaction, is varied from $r = -0.5$ up to -2. The calculated phase diagrams are presented in figures 7a-7c. It is apparent from the graphs that the topology of the phase diagrams presents some clear differences from the corresponding phase diagrams of the systems that lack orientational interactions. Thus, when the strength of the orientational interaction is quite high the nematic phase disappears completely from the phase sequence in favour of the smectic phase (figure 7c). This implies that the directional interactions of the end-blocks strengthen the molecular tendency for microsegregation. A notable consequence of the directional interactions on the layered molecular organisation is the lowering of the degree of interdigitation since molecules of adjacent layers intedigitate only up to the extent that their mesogenic end-blocks are brought to side-by-side register. This is further supported by the fact the Sm$A_{d1}$ is destabilized on increasing the strength of the orientational interactions since Sm$A_{d2}$-like molecular organization allows on average more registered end-blocks per layer compared to the Sm$A_{d1}$-like molecular organization.

Turning now to the twin dendro-mesogenic branch (TDB- $C_{60}$) systems, we show in figures 8a-8d) phase diagrams of pressure *vs* intrinsic probability of the extended antiparallel conformer, for four different values of the interaction parameter $q_0/k_B T$.

The phase diagram in figure 8a has been calculated for $q_0/k_B T = 0.025$, corresponding to weakly repulsive addends. The system exhibits two smectic-*A* phases of which the high-pressure/low-temperature phase is a Sm$A_{d1}$ (fullerene monolayer) and the low pressure is a Sm$A_{d2}$ (fullerene bi-layer). The stability of the phases is due to phase micro-segregation dictated by the molecular partitioning. As seen in figure 8b, increasing the strength of the *inter*-block repulsion to $q_0/k_B T = 0.05$ yields a phase diagram that differs from that of figure 8a in that the Sm$A_{d2}$ phase is stable over a narrower pressure range and a small nematic region appears between the isotropic and Sm$A_{d2}$ regions at high intrinsic probabilities of the extended antiparallel conformer.

For $q_0/k_BT = 0.1$ we have significant changes in the phase diagram, shown in figure 8c, with respect to the phase diagrams obtained with weaker interactions (figures 8a-8b). First, the structure of the low pressure smectic phase is no more inderdigitated: the layer spacing becomes equal to the full length of the extended antiparallel molecular state. Secondly, the interdigitated Sm$A_{d1}$ phase appears only at high pressures and is strongly destabilised at high intrinsic probability of the extended antiparallel conformer. Lastly, the nematic range is significantly broadened. These differences indicate that strengthening the repulsive interactions between the addend blocks enhances the role of the overall molecular shape in driving the molecular self organisation and weakens the influence of sub-molecular partitioning therefore rendering less significant the contribution of the microsegregation mechanism to the molecular ordering. These inferences are further supported by the phase behaviour of the system for $q_0/k_BT = 0.2$. As indicated on the phase diagram in figure 8d, this system does not exhibit any inderdigitated smectic phases. The possible phase sequences are either *I*-Sm*A*, at low intrinsic probability of the extended antiparallel conformer, or *I*-*N*-Sm*A*, at higher probabilities. These are similar to the phase sequences exhibited by sterically interacting rod-like systems [28].

In the limit of very low intrinsic probability for the EA conformer, namely when $P_{EA}^0 \ll 1$ for the TDB model systems or when $\varepsilon_{FP} \ll \varepsilon_{EP}$ for the TMB, the molecules are practically rigid, exhibiting a single conformer with the mesogenic units extended on the same side. Clearly, these molecular shapes correspond to the dominant conformers of the single dendromesogenic branch mono-adducts of $C_{60}$ shown schematically in figure 1c. These systems do not exhibit nematic phases indicating that enhanced molecular polarity disfavours nematic ordering. This is in accordance with what is observed experimentally [7,8,13]. Furthermore, the molecular organisation within the smectic layers corresponds to a bi-layer, "head to tail" arrangement with spacing of about six submolecular blocks. Taking into account that the block length is roughly 9Å (the fullerene diameter), the calculated spacing is found around 55Å, in good agreement with XRD measurements on the real systems [7].

When all the conformers come into play the phase behaviour becomes, richer primarily due to formation of various smectic phases. The polymorphism of the smectic organisation stems from chemical affinity differences between distinct molecular parts in conjunction with molecular flexibility. Thus, the degree of inderdigitation between adjacent layers and the molecular organisation within the layers are determined by the interplay between the molecular flexibility and the formation of well-defined migrosegregated structures. As calculations indicate, phase transitions between smectic phases are accompanied by rather strong conformational changes but not necessarily by substantial changes of the layer spacing. Regarding the smectic phases of the studied TMB and TDB systems, the Sm$A_{d2}$ phases is favoured primarily by the with high probability of the FP conformers while the Sm$A_{d1}$ is more stable when the EA molecular conformers have appreciable intrinsic probability. Both phases exhibit extended intedigitation and their layer spacing differs by one fullerene diameter (9Å).

Finally, for the conical super-mesogens with a fullerene apex (CSM-$C_{60}$), the assumed effective rigidity of the block representation of the molecule removes any dependence of the phase behaviour on the conformational statistics. The block-block interaction potential used for the CSM-$C_{60}$ calculations is formulated and parameterised along the same lines described for TMB-C$_{60}$, and TDB-C$_{60}$ with $u_{f,f}^{(0)} = u_{f,m}^{(0)} = \infty$, $u_{m,m}^{(0)} = q_0 > 0$ and $u_{i,j}^{(1)} = 0$. Here, as before, $q_0$ is a measure of the softness of the repulsions among the grafted addends. Figure 9 shows the calculated phase diagram, the thermodynamic variables in this case being the scaled pressure, $pv_{mol}/q_0$, and the scaled reciprocal temperature $q_0/kT$. It is apparent on that phase diagram that the system, depending on pressure, may exhibit two different phase sequences. At low pressures the system transforms, on lowering the temperature, form the isotropic phase to a columnar phase thought a first order transition. At higher pressures, a bi-layer interdigitated smectic-*A* phase is inserted between the isotropic and the columnar phase.

Strictly, the positional organisation CSM-C$_{60}$ super-mesogens in the plane perpendicular to the columns is forced to a rectangular columnar ordering due to the imposed cubic lattice restrictions on the positions of submolecular blocks. In other words, the axis parallel to the common column orientations can only a C4 or a C2 symmetry axis and therefore the only allowed two dimensional positional order should be consistent with a rectangular symmetry. This makes it impossible to distinguish between hexagonal and rectangular columnar phases within the present lattice model. However, the phase behaviour of the system, in particular the phase boundaries of the columnar phase to the isotropic or to the smectic phase, is not expected to be severely influenced by this limitation since the free-energy difference between a hexagonal and rectangular columnar phases is expected to be rather low compared to the difference between the columnar (rectangular or hexagonal) and nematic or smectic free energy.

The picture for the *intra*-columnar organisation of the CSM-C$_{60}$ super-mesogens is clear: the columns are strongly polar since the molecules stack one on the top of the other so that the fullerene unit of the upper molecule is accommodated inside the cone aperture formed by the addends of the next super-mesogen in the column. It should be noted here that, as a consequence of assuming non-polar block-block interactions, the energy required to slide two adjacent columns parallel to each other does not depend on the their polarity. Accordingly, the overall polarity of these columnar phases is determined solely on entropic grounds, rendering the macroscopically apolar columnar phases more stable than the polar ones.

**IV     Conclusions**
We have studied the phase behaviour and the molecular organization for a wide variety of fullerene containing liquid crystals with the aid a simple molecular theory. Despite the very crude representation of the molecular structure in terms of a small number of sub-molecular blocks, restricted to move on a cubic lattice and interacting via greatly simplified additive block-block potentials, the theory accounts consistently and qualitatively for the basic experimental observations on all the classes of compounds considered. The observed nematic, smectic and columnar phases are reproduced correctly and the molecular features that influence their stability are identified. The peculiar smectic polymorphism exhibited by compounds of twin-branch architecture is elucidated in relation to molecular structure and interactions. While the molecular modeling and the computational aspects of the theory are susceptible to further refinements, the results obtained with its present, simplified, form can be useful for the molecular design of model fullerene containing liquid crystalline compounds.

**Acknowledgments**

SDP acknowledges European Social Fund (ESF), Operational Program for Educational and Vocational Training II (EPEAEK II) and particularly the Program IRAKLEITOS, for funding. AGV acknowledges support through the "Caratheodores" research programme of the University of Patras. Support from the RTN Project "Supermolecular Liquid Crystal Dendrimers – LCDD" (HPRN-CT2000-00016) is also acknowledged.

**List of tables**

**TableI.** Interaction parameters between the building blocks when they occupy the same lattice site $q^{(0)}_{b_1,b_2}$ and when they are in adjacent lattice sites $q^{(1)}_{b_1,b_2}$ (values in brackets).

| $b_2$ \ $b_1$ | $C_{60}$ | Spacer | Mesogens |
|---|---|---|---|
| $C_{60}$ | $\infty$ (0) | $\infty$ ($q_0/2$) | $\infty$ ($q_0/4$) |
| Spacer | $\infty$ ($q_0/2$) | $q_0$ ($q_0/4$) | $q_0/4$ (0) |
| Mesogens | $\infty$ ($q_0/4$) | $q_0/4$ (0) | $q_0/4$ ($q_0/8$) |

# FIGURE CAPTIONS

**Figure 1.** Various architectures of mesogen-functionalized fullerenes: (a)-(c) mono-adducts, (d) bis-adducts, (e) hexa-adducts and (f) conical multi-adducts of $C_{60}$.

**Figure 2.** Space filling and block representations of the dominant conformers of typical mono-adduct $C_{60}$ derivatives. Darker shading in the three top blocks in (c) is used to indicate coincidence of two blocks within a single lattice site.

**Figure 3.** Space filling model (a) and block representations of two representative conformers, (b) and (c), of a typical 2$^{nd}$ generation dendritic mono-adduct $C_{60}$ derivative. Darker shading in all but the bottom block the in (c) is used to indicate coincidence of two blocks within a single lattice site.

**Figure 4** Space filling model (a) and block representation (b) of a multi-adduct fullerene derivative with conical shape. The wire-frame drawing of some blocks has been used in order to provide a better view of the empty space between the grafted addends.

**Figure 5** Calculated phase diagrams (pressure *vs* reciprocal temperature) for the three state inter-converting model of the TMB- $C_{60}$ mono adducts. The EA and L conformers have been taken to have equal intrinsic probabilities ($\varepsilon_{EA} = \varepsilon_L = 0$) and four different cases for the intrinsic probability of the third, FP, conformer were considered: (a) $\varepsilon_{FP}/q_0 = -3$, (b) $\varepsilon_{FP}/q_0 = -1$, (c) $\varepsilon_{FP}/q_0 = 1$ and (d) $\varepsilon_{FP}/q_0 = 3$. The interaction parameter $q_0$ has been set equal to 0.1 in all cases.

**Figure 6.** Calculated bulk probability of the EP (circles), the *L* (triangles) and the FP (squares) conformers, as a function of pressure for the system whose phase diagram is given in figure 5(c) at the fixed value of scaled temperature $q_0/k_B T = 0.2$. Shown on the diagram are also the intrinsic probabilities of the EP and L conformers (solid line) and of the FP conformer (dashed line).

**Figure 7.** Same as in figure 5(b) only with the end blocks (mesogenic units) interacting via an additional directional component of the potential whose strength *r* relative to that of the non-directional component $q_0$ is given by *r*,. (a) $r = -0.5$, (b) $r = -1$ and (c) $r = -2$.

**Figure 8.** Calculated $p^*$, $P_{EA}^0$ phase diagrams (dimensionless pressure vs intrinsic probability of the EA conformer) for the TDB-$C_{60}$ mono-adducts for four different values of the interaction parameter $q_0/k_B T$ : (a) 0.025, (b) 0.05, (c) 0.1 and (d) 0.2 respectively.

**Figure 9.** Calculated phase diagram for CSM-$C_{60}$ conical molecules with a fullerene apex.

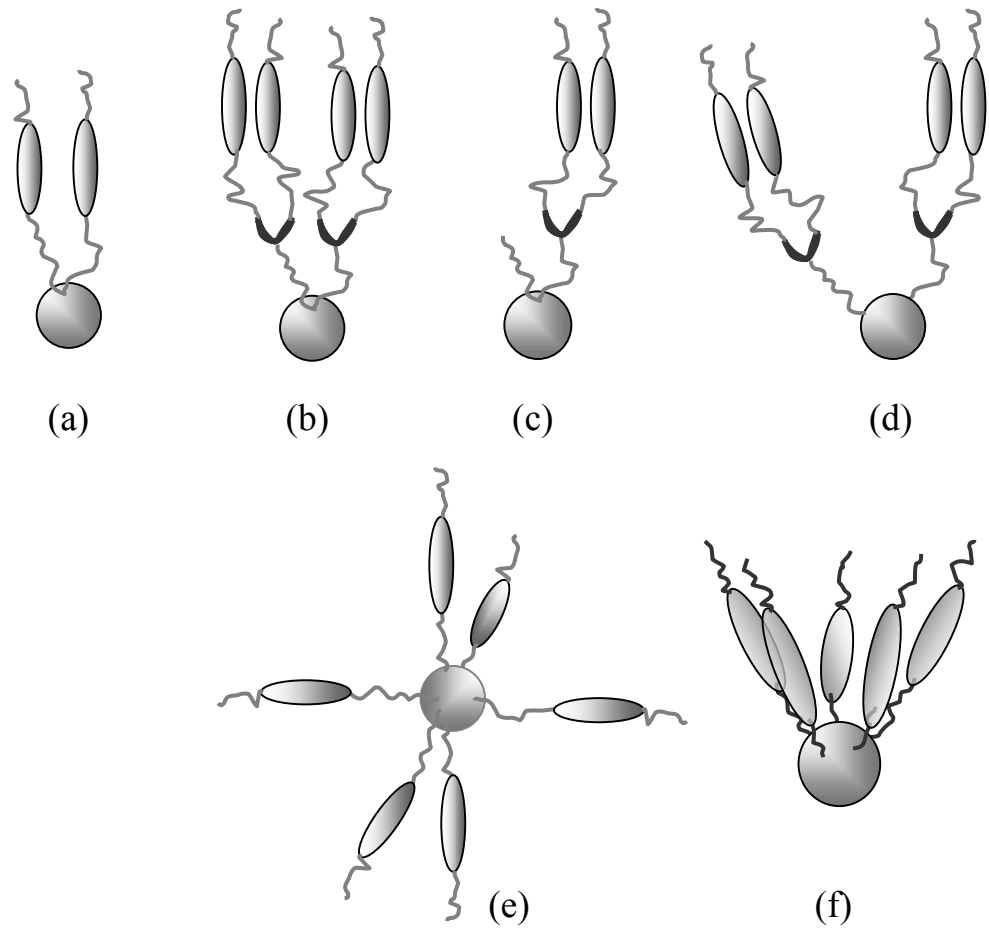

**FIGURE 1**

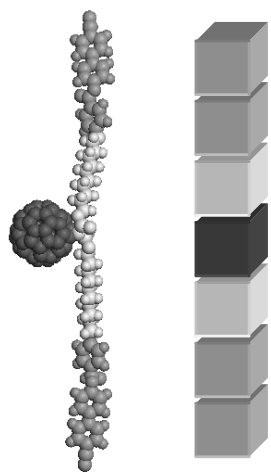 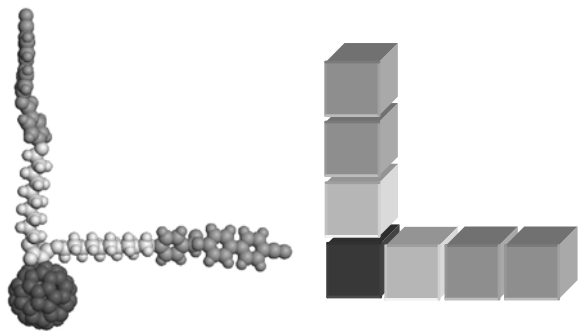 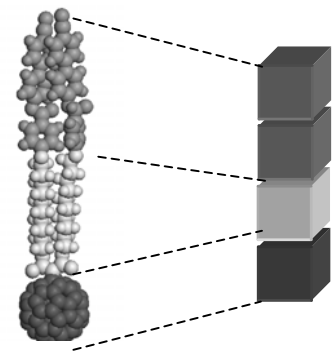

(a) (b) (c)

FIGURE 2

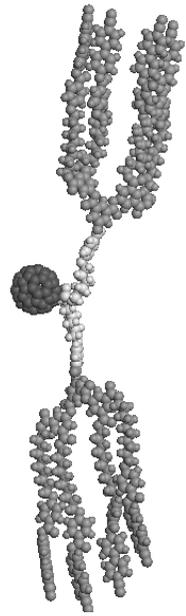 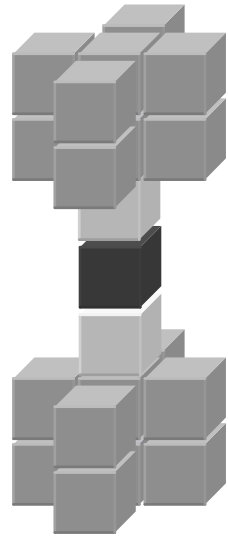 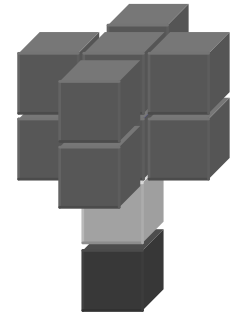

(a)          (b)          (c)

FIGURE 3

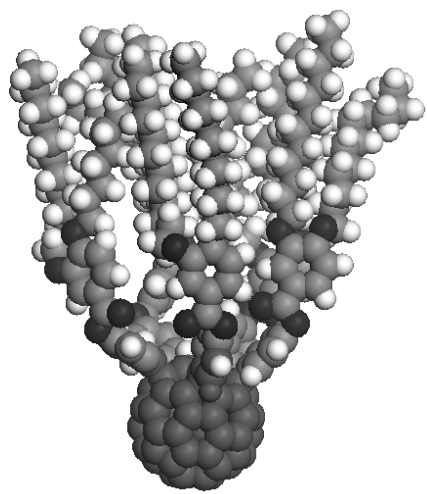 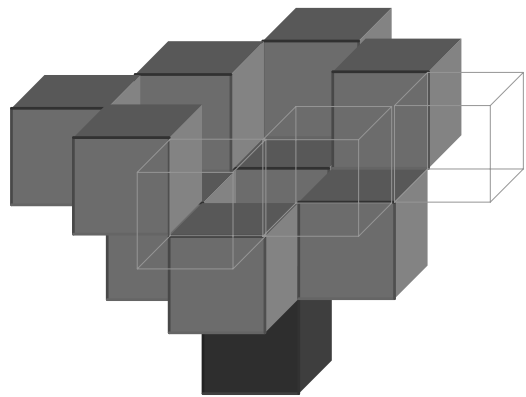

**(a)** **(b)**

FIGURE 4

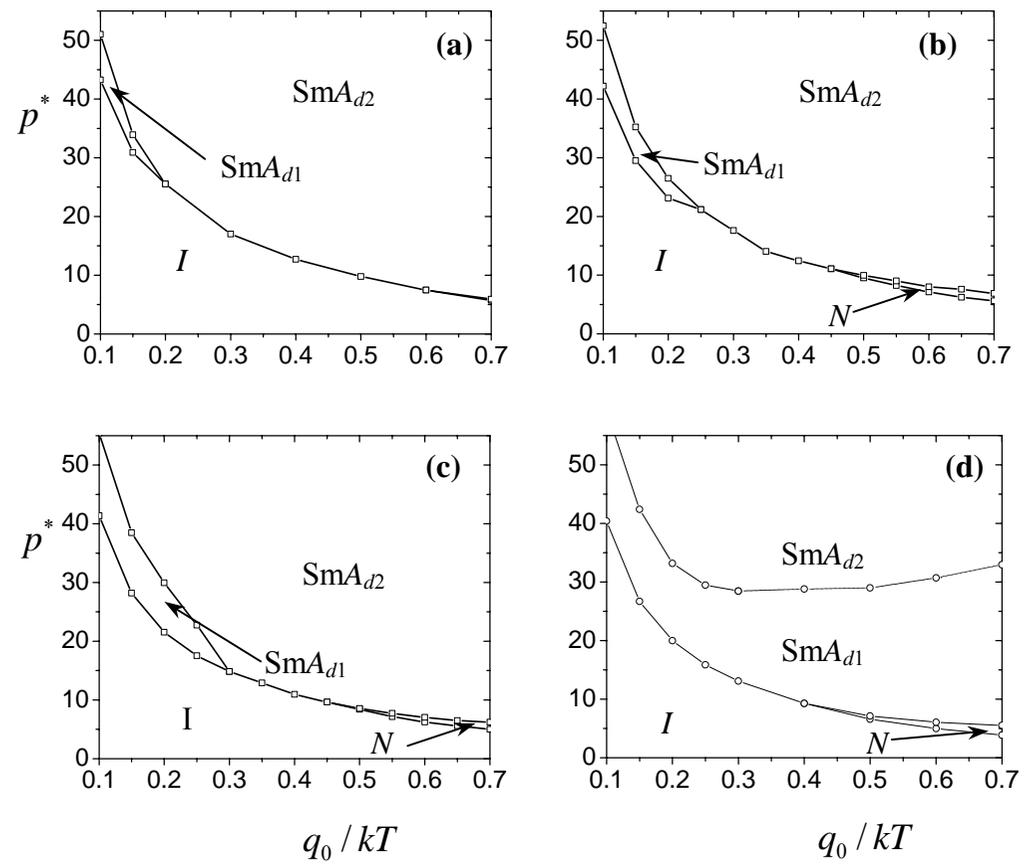

FIGURE 5

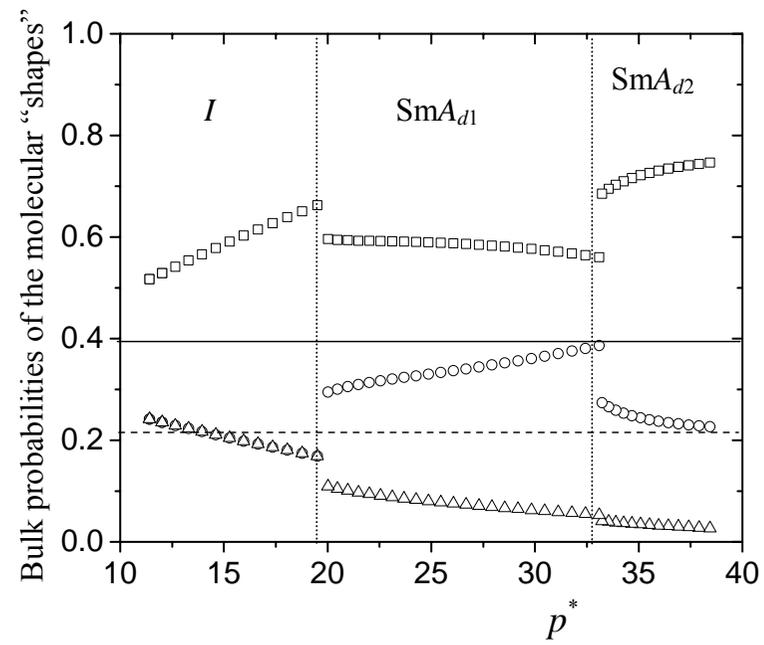

FIGURE 6

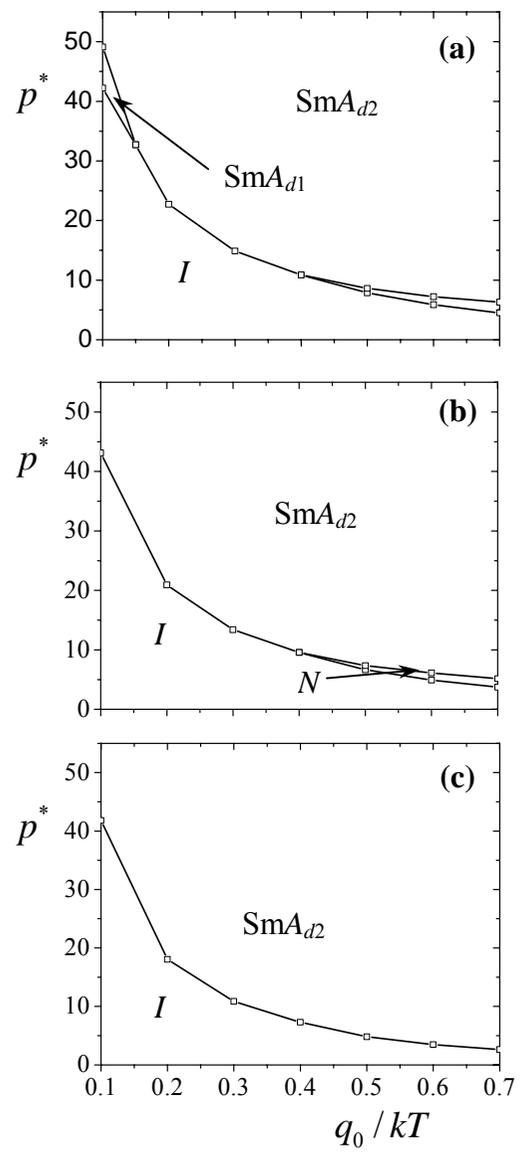

FIGURE 7

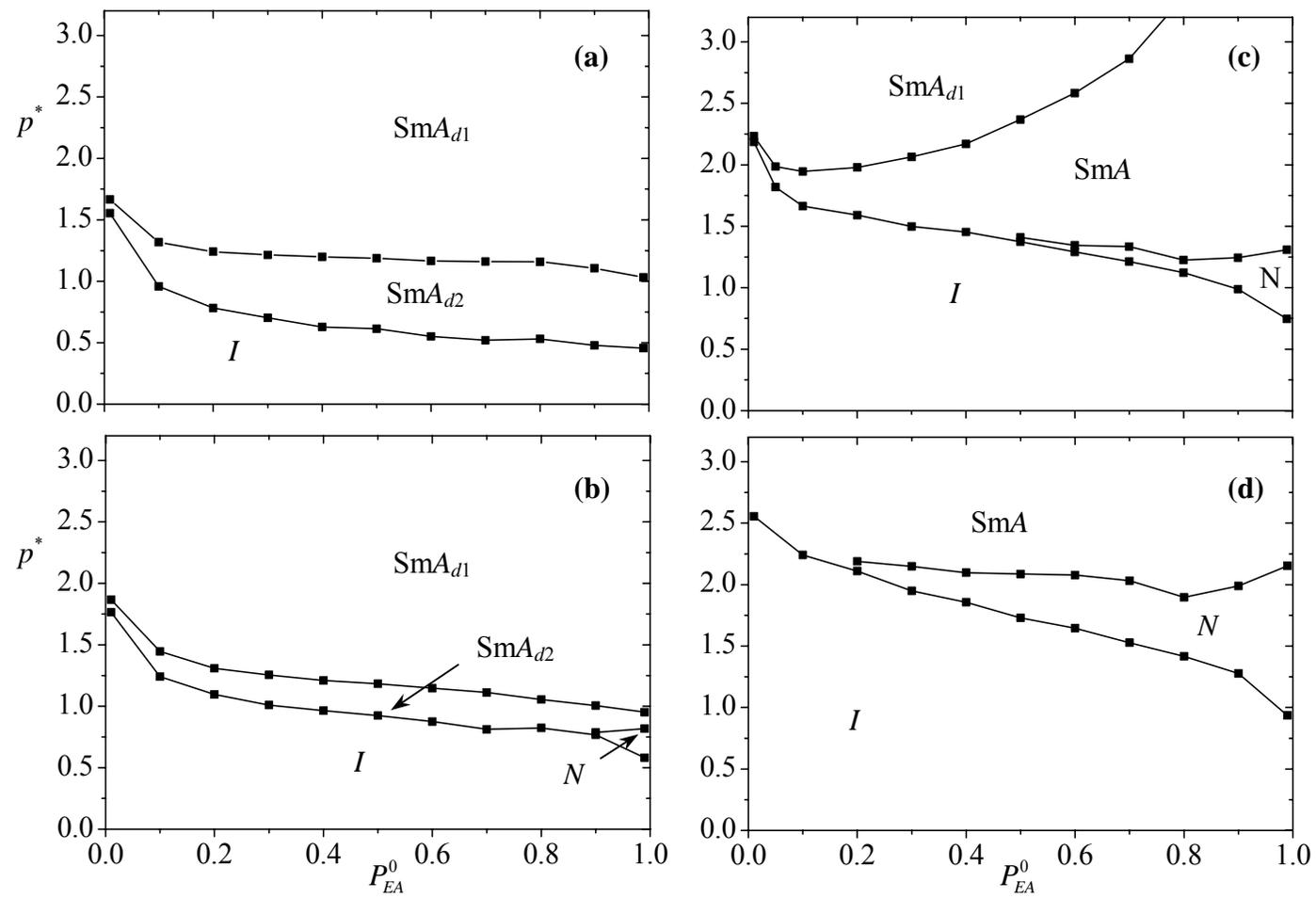

FIGURE 8

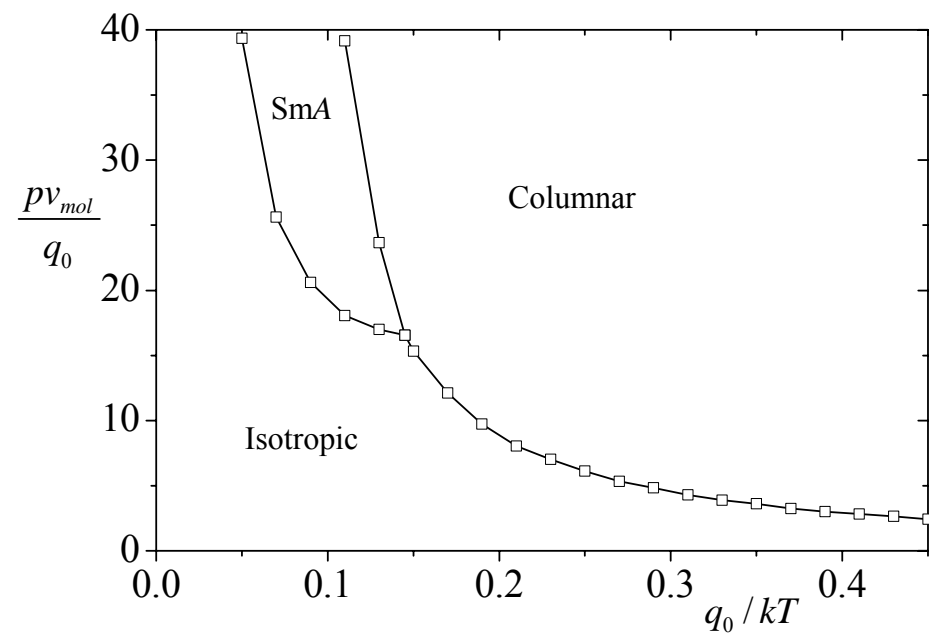

FIGURE 9